\documentclass[aps,prb, twocolumn,groupedaddress]{revtex4-1}
\newcommand{\lb}{\left\langle}
\newcommand{\rb}{\right\rangle}
\usepackage[mathscr]{eucal}
\begin{document}

%\preprint{}

\title{Spin Seebeck effect and magnon-magnon drag in Pt/YIG/Pt structures}


\author{ I.I. Lyapilin$^{1,2}$, M.S. Okorokov$^1$}
\email {Okorokovmike@gmail.com}
%\thanks{}
\affiliation{$^1$Institute of Metal Physics, UD RAS, Ekaterinburg, 620137, Russia\\
$^2$Ural Federal University after the first President of Russia B.N. Yeltsin, Yekaterinburg, 620002 Russia}


\date{\today}

\begin{abstract}
The formation of the two: injected (УcoherentФ) and УthermallyФ excited, different in energies magnon subsystems and the influence of its interaction with phonons and between on drag effect under spin Seebeck effect conditions in the magnetic insulator part of the metal/ferromagnetic insulator/metal structure is studied.  An approximation of the effective parameters, when each of the interacting subsystems ("injected"\,, "thermal"\, magnons, and phonons) is characterized by its own effective temperature and drift velocities have been considered. The analysis of the macroscopic momentum balance equations of the systems of interest conducted for different ratios of the drift velocities of the magnon and phonon currents show that the УinjectedФ\, magnons relaxation on the УthermalФ\, ones is possible to be dominant over its relaxation on  phonons.  This interaction will be the defining in the forming of the temperature dependence of the spin-wave current under spin Seebeck effect conditions, and inelastic part of the magnon-magnon interaction is the dominant spin relaxation mechanism.
\end{abstract}
\pacs{72.15 -b,  71.15}
\keywords{Magnons; phonons; drag effect; spin current; exchange interaction; thermoelectric coefficients}

\maketitle


\section{Introduction}

The influence of non-equilibrium phonons on kinetic coefficients in electron-phonon or spin-phonon systems has been theoretically studied chiefly by the Boltzmann kinetic equation method or using the formalism of the Kubo response theory. The present work employs the method of the non-equilibrium statistical operator (NSO) to analyze how interactions between three flows (two magnon and phonon ones) affect the drag effect and the temperature dependence of the spin Seebeck effect (SSE) within the above model.

The concept of magnon spintronics, i.e., the  generation,  detection   and manipulation of pure spin currents in the form of spin wave quanta  \cite{1}, the magnons, has attracted growing interest in the recent years \cite{01, 02}.  Magnons are quasiparticles representing a low-energy excited state of ferromagnets. A quantized magnon is a boson and carries basic spin angular momentum quanta of $\hbar$ \cite{2}. Similar to spintronics and electronics, magnonics refers to using magnons for data storage and information processing \cite{3}. Up to now, magnons in the field of spintronics have been investigated within the content of magnetostatic spin  waves which describe the nonuniform spatial and temporal distribution of the classical magnetization vector.

Recently, a  great deal  of attention  is devoted  to the  investigation of  thermally  excited magnons,  particularly in studies of the spin Seebeck effect \cite{3, 4, 5, 6, 7} in Pt/YIG/Pt structure. In SSE effect, an applied electric current in one Pt layer accompanies an electron spin current due to the spin Hall effect (SHE) \cite{6, 7}. When the spin current flows to the boundary between the Pt and  the YIG, nonequilibrium spins are accumulated and, consequently, due to the s-d exchange interaction between conduction electrons in normal metal (NM) and magnetic moments in ferromagnetic insulator (FI), magnons are created at the interface \cite{8, 9}. The induced magnons subsequently diffuse in FI to the other interface where the magnon current converts back to an
electron spin current in the other NM layer, leading to a charge current due to the inverse spin Hall effect (ISHE) \cite{10, 11, 12}. Thus, the induced electric current in the second NM layer which is electrically insulated from the current-flowing NM layer by a FI would elucidate the magnons as spin information carriers.  One of the key advantages of magnon  spin currents is their large damping length, which can be several orders of magnitude higher than the spin diffusion length in conventional spintronic devices based on spin-polarized electron currents \cite{13}.

The propagation of magnons in a magnetic insulator is described by two characteristic quantities: mean free path and spin diffusion length that are governed, in turn, by various magnon relaxation mechanisms. A series of experiments determine the range of the diffusion lengths as being quite wide: from 4$\mu$m to 120$\mu$m \cite{14, 15, 16, 17, 18}. To explain so large values of the spin diffusion lengths, the number of papers has put forward several concepts of
appearance, along with "thermal"\, magnons, of long-wave ("subthermal"\,) magnons in a magnetic insulator. Under SSE conditions, the former are characterized by short wavelength, the latter are weakly coupled with the lattice \cite{15, 18}. For interpreting the experimental results, the works \cite{15, 18, 19} have adopted the hypothesis of the existence of the two magnon subsystems with different energies. As to the temperature dependence of the Seebeck coefficient, it is non-monotonic and reaches its maximum within the range of 50 - 100K. And as the investigations have shown, it is affected by strength of a magnetic field, dimensions of the samples, and quality of the interface \cite{18, 20}.

To explain the low temperature enhancement \cite{21} proposed the phonon-drag SSE scenario based on a theoretical model \cite{22, 23, 24}.   Back in 1946, in the context of thermoelectricity, Gurevich pointed out that thermopower can be generated by nonequilibrium phonons driven by a temperature gradient, which then drag electrons and cause their motions \cite{22}.   It was suggested first by Bailyn \cite{25} that the theory of magnon-drag should be analogous to that of  phonon-drag \cite{22}. Following this work the magnon-drag component of the thermopower has been calculated by Grannemann and Berger \cite{26}. In this phenomenological model, the temperature dependence of the phonon life time is involved, which reaches a maximum at low temperatures. Based on this, they propose a strong interaction between the phonons and magnons, which are responsible for the heat transport in the system. The phonons flow along the thermal gradient and interact with thermally excited magnons. These phonons УdragФ the magnons. Thus, the phonon-magnon coupling is suggested to explain the observed enhancement of SSE signal at low temperature \cite{21, 22}.  The first investigation of the phonon-magnon interaction in magnetic insulators was conducted by Adachi et al., reporting a giant enhancement of SSE in $ LaY_2Fe_5O_{12}$ at low temperatures \cite{21}. However,  the observed transport of magnons over a long distance of up to millimeters in magnetic insulators implies a relative weak interaction with phonons and impurities, and the  measurements of the temperature dependent thermal conductance of YIG single crystals show that the phonon contribution to the thermal conductivity reaches its maximum at around 25 K \cite{14}, which is ~50 K lower than the observed peak in the SSE.  In \cite{26} the temperature dependence of the spin Seebeck effect was measured in $(Ga,Mn)As$, and the data showed a pronounced peak at low temperatures.

Here, we study how the formation of two interacting magnon subsystems with distinguished energies affects the SSE \cite{15, 17, 18, 19}. We assume that the first group of magnons is "thermal"\, ones subjected to a non-uniform temperature field applied to the magnetic insulator. The energy of such magnons is of the order of the temperature $k_BT$. Further, under the SSE, inelastic scattering of spin-polarized electrons of the metal by localized spins located near the interface causes the magnons to inject into the magnetic insulator. The energy of the injected ("coherent") magnons is of the order of the spin accumulation energy $\Delta_s$ of conduction electrons of the metal. It can be generated, for example, by the spin Hall effect when passing a direct electric current through the metal (Pt) [1].

Under the SSE conditions, the injection of magnons into the magnetic insulator dominates scattering processes with magnon absorption provided that the inequality  $\Delta_s>k_BT$ is fulfilled. Thus, it can be said that the magnetic system of the insulator forms another subsystem of  "injected"\,("coherent")\, magnons that are actually responsible for the SSE. As a consequence, in the  presence of a non-uniform temperature field, there are three flows inside the magnetic insulator, namely, phonon and two magnon ones. The evolution of the magnon and phonon subsystems to equilibrium occurs due to the relaxation of both their energy and their moment. These subsystems tend to become balanced with different velocities. Obviously, the interaction between the flows gives rise to the drag effect \cite{27, 28, 29}.

It is worth noting that the paper \cite{30} has already discussed the influence of mutual dragging between phonons and spin excitations on thermal conductivity of a spin system. Once the magnon subsystems are thermalized in energy, the magnon system can be characterized solely by a temperature $T_m$. As to the moment relaxation processes, this time is defined, as well as in the event of spin-flip electron scattering, by inelastic magnon scattering. The time relaxation is large enough, which, apparently, provides the existence of the SSE over long distances \cite{14}.

The paper is structured in the following manner. The first part is devoted to the splitting of the "injected"\, magnons flow responsible for the spin Seebeck effect from the magnon current. In the second part the balance equations for the magnons and phonons in the approximation of effective parameters when each subsystem is characterized by its effective temperature and  drift velocity are built and analysed.

\section*{Spin current}

The density of the spin current $J_s({\bf r})$  can be represented as the sum of two terms: collisional $\dot{s}^z_{(sm)}({\bf r})$ and collisionless $I_{s^z}({\bf r})$. The former is controlled by the inelastic spin-flip electron scattering by localized moments at the interface, the latter is due to the flows of electrons with different spin orientation \cite{31}
\begin{eqnarray}\label{1}
J_s({\bf r})=\frac{d}{dt}s^z({\bf r}) &=& \frac{1}{i\hbar}\,[s^z({\bf r}), H\,] =
-\nabla\,I_{s^z}({\bf r})+ \dot{s}^z_{(sm)}({\bf r}),\nonumber\\
    I_{s^z}({\bf r}) &=& \sum_i\,s^z_i\{p_i/m, \delta({\bf r}-{\bf r}_i)\}, \nonumber\\
    \dot{s}^z_{(sm)}({\bf r}) &=& \frac{1}{i\hbar}\,[s^z({\bf r}), H_{sm}\,].
\end{eqnarray}
Here $H$ is the Hamiltonian of the system considered, $H_{sm}$  is the density of the exchange interaction energy between conduction electrons and localized moments at the interface \cite{32}.
\begin{equation}\label{2}
    H_{sm} = -J_0\!\sum_j\int\!\!d{\bf r} {\bf s}({\bf r})\,{\bf S}({\bf R}_j)\,\delta({\bf r}-{\bf R}_j),
\end{equation}
$J_0$  is the exchange integral, ${\bf S}({\bf R}_j)$  is the operator of a localized spin with the coordinate ${\bf R}_j$  at the interface. ${\bf s}({\bf r})$ is the spin density of the electrons in the metal. Accompanied by the creation of magnons, the inelastic scattering of spin-polarized conduction electrons by localized impurity centers dominates other processes. This leads to magnon accumulation  $(\delta N({\bf r}) = N({\bf r}) - N^0({\bf r})\,)$\cite{31}, $N({\bf r}),\, N^0({\bf r})$ are the non-equilibrium and equilibrium magnon distribution functions near the interface in the magnetic insulator. The macroscopic spin current can be found by averaging the expression with the non-equilibrium statistic operator $\rho(t)$ \cite{33}:
\begin{equation}\label{2a}
\lb J_s({\bf r})\rb^t = - \nabla\,\lb I_{s^z}({\bf r})\rb^t + \lb\dot{s}^z_{(ms)}({\bf
r})\rb^t,
\end{equation}
where $\lb\ldots\rb^t = Sp (\rho(t)\ldots)$.

Given that $[s^z_i, s^\pm_j] = \pm s^\pm_i\delta_{ij}$, we have
 \begin{eqnarray}
  \lb\dot{s}^z_{(ms)}({\bf r})\rb^t = (-J_0/2)\,\sum_j\int\,d{\bf r} \lb ( s^+({\bf r}) S^-({\bf R}_j)-\right.\nonumber\\
\left.-  s^-({\bf r})\, S^+({\bf R}_j))\,\delta({\bf r}-{\bf R}_j)\rb^t\qquad
\end{eqnarray}

Restricting ourselves to the linear approximation in the interaction $H_{sm}$, we omit the interaction $H_{sm}$  in the operator $\rho(t)$   in which the averaging is performed. In this case $<\ldots>^t = <\ldots>_e^t\,<\ldots>_m^t$, i.e. the averaging in the electron system and the localized spin system is carried out separately:
 $$\lb s^\alpha({\bf r})\,S^\beta({\bf R}_j)\rb^t \rightarrow \lb s^\alpha({\bf r})\rb^t_e\,\lb S^\beta({\bf R}_j)\rb^t_m.$$
Thus, we arrive at,
 \begin{eqnarray}\label{2b}
 \lb\dot{s}^z_{(ms)}({\bf r})\rb^t = (-J_0/2)\sum_j\int\,d{\bf r}\{\lb  s^+({\bf r})\rb^t_e\,\lb S^-({\bf R}_j)\rb^t_m -\nonumber\\
 - \lb s^-({\bf r})\rb^t_e\,\lb S^+({\bf R}_j)\rb^t_m\}\,\delta({\bf r}-{\bf R}_j).\qquad\qquad
\end{eqnarray}
Let us calculate  $S^\pm({\bf R})$. We write down the equations of motion for the transverse components
$$\dot{S}^\pm({\bf R}) = (i\hbar)^{-1} [ S^\pm({\bf R}),\, H_m  + H^k_m + H_{mp} + H_{ms} ],$$
where $ H_m,\, H^k_m,\, H_{mp},\, H_{ms}$   are the energy  operators of the magnetic subsystem (Zeeman and kinetic), the magnon-phonon $(mp)$ and exchange $(sm)$ interaction, respectively. Computing the commutators, we obtain
\begin{eqnarray}\label{3}
\dot{S}^\pm({\bf R}) = \mp i\omega_m S^\pm({\bf R}) -\nabla I_{S^\pm}({\bf R}) +  \dot{S}^\pm_{(mp)}({\bf R}) + \dot{S}^\pm_{(ms)}({\bf R}),\nonumber\\
\dot{A}_{(ik)}({\bf R})= (i\hbar)^{-1}[A({\bf R})\,, H_{ik}]\qquad\qquad\qquad
\end{eqnarray}
and the density of the magnon flows at the interface
\begin{equation}\label{4}
    I_{S^\pm}({\bf R}) =\sum_j\,S^\pm_j\{P_j/M,\delta({\bf R} - {\bf R}_j)\}
\end{equation}
$P_j, M$ are the magnon momentum and the effective magnon mass. The last two summands in the right-hand side of (\ref{3})describe the scattering of the magnons by phonons and electrons at the interface. Conducting the averaging, in the stationary case we come to
\begin{eqnarray}\label{5}
\mp i\omega_m \lb S^\pm({\bf R})\rb_m  =  \nabla \lb I_{S^\pm}({\bf R})\rb_m -\nonumber\\
- \lb\dot{S}^\pm_{(mp)}({\bf R})\rb_m  - \lb\dot{S}^\pm_{(ms)}({\bf R})\rb_m.
\end{eqnarray}
Further, we insert the expression (\ref{5}) into the equation for the spin current $\lb\dot{s}^z_{(ms)}({\bf r})\rb$ and estimate the summands.  The first term in the right-hand side of the expression (\ref{2b}), approximately proportional to $\sim J_0$, governs the magnon flow excited at the interface due to electron scattering by localized moments. The second term is proportional to $\sim J_0U_p$  and sets forth the magnon scattering by phonons ($U_p$ characterizes the intensity of the magnon-phonon interaction). Finally, the last term in the right-hand side of the expression (\ref{2b}) $\sim J_0^2$. Putting that $U_p\gg J_0$, we leave this term aside.

Let us unravel the evolution of the magnetic subsystem.
$$\dot{S}^z({\bf R}) = (i\hbar)^{-1} [ S^z({\bf R}),\, H_m + H^k_m + H_{(mp)} + H_{(ms)} ]$$
Then we have
\begin{equation}\label{6}
\dot{S}^z({\bf R}) = -\nabla  I_{S^z}({\bf R}) + \dot{S}^z_{(ms)}({\bf R}) +
\dot{S}^z_{(mp)}({\bf R}).
\end{equation}
The first term in the right-hand side of (\ref{6}) involves the spin-density flow of localized spins (magnons), and the terms $\dot{S}^z_{(mp)}({\bf R}) ,\,\dot{S}^z_{(ms)}({\bf R}) $, described the scattering of the localized spins by phonons at the interface.

Thus, the macroscopic spin-wave current realized in the magnetic insulator can be written as
\begin{eqnarray}\label{8}
\lb I_S({\bf R})\rb =   -\nabla\lb  I_{S^z}({\bf R})\rb + \lb\dot{S}^z_{(ms)}({\bf R})\rb_m + \lb\dot{S}^z_{(mp)}({\bf R})\rb_m+\nonumber\\
 +(-J_0)\,\sum_j\int d{\bf r} \{ \lb s^+({\bf r})\rb_e\,\lb {\bf S}^- ({\bf R}_j)\rb_m)- \nonumber\\
 -\lb s^-({\bf r})\rb_e\,\lb{\bf S}^+({\bf R}_j)\rb_m\}\,\delta({\bf r}-{\bf R}_j),\qquad\qquad
\end{eqnarray}
where
\begin{equation}\label{9}
\lb S^\pm({\bf R})\rb_m\!\!  =\! i\omega_m^{-1}\{\! - \nabla\!\lb I_{S^\pm}({\bf R})\rb_m\!\! -\! \lb\dot{S}^\pm_{(mp)}({\bf R})\rb_m\!\big\}.
\end{equation}
In (\ref{9}), we have omitted the summand that describes the magnon scattering at the interface ( $\sim J_0^2$). It can be seen from (\ref{8}), (\ref{9}) that the magnetic subsystem realizes two magnon flows. The first is  due to a non-uniform temperature perturbation of the magnetic subsystem. It is a flow of "thermal"\, magnons. The mean energy of these magnons is of the temperature. The second is $\sim J_0\,\nabla I_{S^\pm}({\bf R})$  and is brought about by magnons injected into the magnetic subsystem as a result of inelastic scattering of conduction electrons by localized moments at the interface. The energy of such magnons is of the order of the spin-accumulation energy of the conduction electrons and is equal to $\Delta_s\gg k_bT$.

\section*{Macroscopic momentum balance equations }

The influence of non-equilibrium phonons on kinetic coefficients in electron-phonon or spin-phonon systems has been theoretically studied chiefly by the Boltzmann kinetic equation method or using the formalism of the Kubo response theory \cite{30}. The present work employs the method of the non-equilibrium statistical operator (NSO) for analyzing how interactions between three flows (two magnon and phonon one) affect the drag effect and the temperature dependence of the spin Seebeck  effect within the above model. In constructing macroscopic momentum balance equations for the system at hand, we should use the Hamiltonian
\begin{equation}\label{d1}
    H =  H_M + H_P + H_V
\end{equation}
Here $H_M$  is the Hamiltonian of the magnetic system that consists of two magnetic subsystems: of\, "injected"\,("coherent")\, $(H_{m_1})$  and "thermal"\,$(H_{m_2})$ magnons and their mutual interaction
\begin{equation}\label{d2}
H_M =  \int d{\bf r}\, (\sum_iH_{m_i}({\bf r}) +  H_{m_im_i}({\bf r}) ),\quad i=1, 2
\end{equation}
 The integration is performed over the volume occupied by the magnetic insulator FI.
$H_{m_i}({\bf r}) $ is the energy density operator of the (i) magnetic subsystem.  $ H_{m_im_i}({\bf r})$ is the Hamiltonian of the magnon-magnon interaction  inside each the subsystems

Suppose the magnon gas to be free:  $H_{m_im_i} = \sum_{\bf k} \varepsilon({\bf k}) b^+_{\bf k}\,b_{\bf k},\,\, \,\mbox{где}\,\,\, \varepsilon({\bf k}) = P^2/(2 M)$ is the sum of the energies of quasi-particles, ferromagnons having a quasi-momentum $ {\bf P}=\hbar {\bf k}$ with their effective mass $M$ and magnetic momentum \cite{32}). $b^+_{\bf k},\,b_{\bf k}$ are the creation and annihilation operators for the magnons with the wave vector ${\bf k}$.

$H_p$  is the lattice Hamiltonian
\begin{equation}\label{d3}
H_p = \int d{\bf r} \,(H_p({\bf r}) +H_{pp}(\bf r)),
\end{equation}
where $H_p({\bf r})$ is the energy density operator for the phonon subsystem. $H_{pp}({\bf r})$ is the phonon scattering by non-magnon relaxation mechanisms (scattering by the boundaries of the sample, impurities and defects of the lattice, etc.)
\begin{equation}\label{d4}
H_V({\bf r}) =H_{m_im_j}({\bf r}) + H_{m_ip}({\bf r})+  H_{m_is}({\bf r})
\end{equation}
is the energy density operator of interaction between the phonons. $H_{m_ip}({\bf r})$ is the energy density operator of interaction between the phonon and magnetic (i) subsystems. $H_{m_im_j}({\bf r})$ describes the interaction between "thermal"\, and \,"coherent"\, (injected)\, magnons.
$ H_{m_is}({\bf r})$ is the energy density operator of exchange interaction between conduction electrons  and localized magnetic moments at the interface.

Under the influence of a non-uniform temperature field (a temperature gradient) applied to the system, the magnons and phonons begin travelling; their macroscopic drift affects the propagation of the spin-wave current. Obviously, the drag effects that may arise in the system considered are governed by both magnon-phonon collision frequencies and phonon relaxation mechanisms by other mechanisms of their scattering. The problem to be solved reduces to constructing and analyzing a set of macroscopic momentum balance equations $\dot{P}_i({\bf r}) = (i\hbar)^{-1} [P_i({\bf r}), H]\,\  $
 for the magnon ($ i = 1, 2$ ) and phonon ($ i = p$ ) subsystems.

In writing the Hamiltonian, we have omitted the exchange interaction between localized spins and conduction electrons at the interface. In doing so, we have put that it is the exchange interaction that is responsible for the magnon injection into the magnetic insulator and makes no significant contribution to the momentum relaxation of the magnons and phonons.

The equations of motion for the magnon and phonon momenta have the form:
 \begin{eqnarray}\label {d14}
    \frac{\partial }{\partial t}P_{m_i} ({\bf r})& =&  - \nabla I_{P_{m_i}}({\bf r})  + \dot{P}_{(m_i,v)} ({\bf r}),\quad (i\neq j  = 1, 2)\nonumber\\
    \frac{\partial }{\partial t}P_p({\bf r}) & = & - \nabla I_{P_p}({\bf r}) + \dot{P}_{(p,pp)}({\bf r}) + \dot{P}_{(p,v)}({\bf r}),
\end{eqnarray}
where
$$ \dot{A}_{(i,v)} = (i\hbar)^{-1} [A_i, H_{v}].$$
The first terms in the right-hand sides of (\ref{d14}) are the flows of appropriate momenta $I_{P}({\bf r}) = \sum_i\{P_i/M, \delta({\bf r} - {\bf r}_i)\}$. The rest of the terms in the right-hand side of these equations describe the relaxation processes: magnon-phonon and magnon-magnon scattering.

To derive the macroscopic equations (\ref{d14})   $$\lb \dot{P}_i({\bf r})\rb^t= Sp\{ \dot{P}_i({\bf r})\,\rho (t)\},\, (i = m_1, m_2, p),$$ the expression for the NSO needs to be sought. According to \cite{33, 34}, for $\rho(t)$  we have:
\begin{eqnarray}\label {d15}
\rho(t) = \epsilon\int\limits_{-\infty}^0 dt' \,e^{i\epsilon t'}\,e^{it'L}\rho_q(t+t'), \qquad \epsilon\rightarrow +0,\nonumber\\
\rho_q(t) = e^{-S(t)},\quad e^{itL}A = e^{-itH/\hbar}\,A\,e^{itH/\hbar},
\end{eqnarray}
Here $\rho_q(t)$  is the quasi-equilibrium statistical operator. The non-equilibrium state of the system considered corresponds in terms of average density values to the entropy operator
\begin{eqnarray}\label{d16}
S(t) = S_0+\! \delta S(t) =\qquad\qquad\nonumber\\
=\Phi(t)\! +\!\int\! d{\bf r}\{\beta_{m_i} ({\bf r}, t)[H_{m_i}({\bf r}) \!+ H_{m_im_i}({\bf r})+\nonumber\\
 + H_{m_im_j}({\bf r})] \!- \!\beta\mu_{m_i}({\bf r}, t) N_{m_i}({\bf r}) +\nonumber\\
 + \beta_p({\bf r}, t) [H_p({\bf r})\! +H_{pp}({\bf r})\! + H_{pm_i}({\bf r})]  -\nonumber\\
  -\beta_{m_i} ({\bf r},t)V_{m_i} ({\bf r}, t) P_{m_i}({\bf r})\! +\beta_p ({\bf r}, t)V_p ({\bf r},t) P_p({\bf r})\}.\qquad
\end{eqnarray}
Here $S_0$  is the entropy operator for the equilibrium system. $\delta S(t)$ describes the deviation of the system from its equilibrium state. $\Phi(t)$  is the Massieu-Plank functional. $\beta_{m_i}({\bf r}, t)$ are local-equilibrium values of the inverse temperatures of the magnon $(i=1,2)$ and phonon subsystems $\beta_p({\bf r},t)$. $\mu_{m_i}({\bf r}, t)$is a local equilibrium value of the  chemical potential of the magnons. $N({\bf r})=N_{m_1}({\bf r}) + N_{m_2}({\bf r})$  is the magnon number density operator. $V_{m_i}$, $V_p$ are the drift velocities of the magnons $(i = 1, 2)$ and the phonons respectively.
%-------------------------------------------------------------------------------------------------

Magnons, as well as phonons are Bose particles; their distribution function is the Bose-Einstein function with a zero chemical potential. However, the situation becomes quite different if magnons are non-equilibrium. In our case, the non-equilibrium magnon system may be described by introducing the non-equilibrium chemical potential of magnons \cite{35, 36, 37}.

 For $\rho(t)$, in the linear approximation in deviation from equilibrium, we arrive at
   \begin{equation}\label{d17}
\rho(t) =\rho_q(t) - \int\limits_{-\infty}^0\,dt'\,e^{\epsilon t'}e^{it'L}\int\limits_0^1d\tau\, \rho_0^\tau\dot{S}(t+t')\rho_0^{-\tau}\,\rho_0.
\end{equation}
 $\dot{S}(t)= \partial S(t)/\partial t +(i\hbar)^{-1}[S(t), H]$ is the entropy production operator. $\rho_0=\exp\{-S_0\}$.Thus, the problem boils down to finding the entropy production operator.

We write down the equations of motion for the operators involved in the entropy operator. Then, we have
 %--------------------------------------------------------------------------------------------------
 \begin{eqnarray}\label {d18}
         \dot {H}_{m_i}({\bf r})& =& - \nabla I_{H_{m_i}}({\bf r}) +\dot{H}_{(m_i,v)}({\bf r}) \nonumber\\
         \dot{H}_p({\bf r})   &= &- \nabla I_{H_p}({\bf r}) +\dot{H}_{(p,pp)}({\bf r}) +\dot{H}_{(p,v)}({\bf r}),\nonumber\\
          \dot{N}_{m_i}({\bf r}) &= & - \nabla I_{N_{m_i}}({\bf r}) + \dot{N}_{(m_i,v)}({\bf r}) .
\end{eqnarray}
The first terms in the right-hand sides of these equations control the flows of appropriate quantities: the energy and number of magnons, phonons, meanwhile, the rest of the terms describe relaxation processes. $I_{N_m}({\bf r})$ is the density of the magnon flow.

 Substituting the equations of motion into the entropy production operator, we come to
 %-----------------------------------------------------------------------------------------------------------------------------
 \begin{eqnarray}\label{d19}
\delta\dot{S}(t)\!\! =\! \!\Delta\!\!\! \int\!\! d{\bf r}\{\!-\delta\beta_{m_i}({\bf r},t)\nabla I_{H_{m_i}}({\bf r})\! +\qquad\qquad\nonumber\\
+\! \beta\mu_{m_i}({\bf r},t)\nabla I_{N_{m_i}}({\bf r})-\delta\beta_p({\bf r},t)\nabla I_{H_p}({\bf r})\!+\qquad\qquad\nonumber\\
+ \!\beta_{m_i}({\bf r},t) V_{m_i}({\bf r},t)\nabla I_{P_{m_i}}({\bf r})\! +\beta_p({\bf r},t) V_p({\bf r},t) \nabla I_{P_p}({\bf r})+
\qquad\nonumber\\
+\delta\beta_{m_im_j}({\bf r},t)\dot{H}_{(m_i,m_im_j)}({\bf r})
-\beta\mu_{m_i}({\bf r},t)\dot{N}_{(m_i,v)}({\bf r})-\qquad\nonumber\\
\beta_{m_i}({\bf r},t) V_{m_i}({\bf r},t) \dot{P}_{(m_i,v)}({\bf r})-\beta_p({\bf r},t) V_p({\bf r},t)\dot{P}_{(p,v)}({\bf r})\},\qquad
\end{eqnarray}
 where $ \delta\beta_{m_im_j}\! =\!\beta_{m_i}\! -\!\beta_{m_j},\,\,\,\Delta A\! =\! A -\! <A>_0.$
 %------------------------------------------------------------------------------------------------------------------------------

We integrate by parts the terms containing the flow divergences. Then, we ignore the surface integrals and write down the entropy operator as
\begin{eqnarray}\label{d22}
\dot{S}(t)\!=\!\Delta\!\int\! d{\bf r}\{-\beta\,I_{N_{m_i}}({\bf r})\nabla\mu_{m_i}({\bf r}, t) +\qquad\nonumber\\ +\delta\beta_{m_im_j}({\bf r},t)\dot{H}_{(m_i,m_im_j)}({\bf r})\! +\!\,I^*_{m_i}({\bf r})\nabla\beta_{m_i}({\bf r}, t) +\nonumber\\
+\,I^*_p({\bf r})\,\nabla\beta_p({\bf r}, t)\!-\!\beta\mu_{m_i}({\bf r},t)\dot{N}_{(m_i,v)}({\bf r}) -\nonumber\\
- \beta V_{m_i}({\bf r}, t)\dot{P}_{(m_i,v)}({\bf r})\! -\! \beta V_p({\bf r},t)[\dot{P}_{(p,pp)}\! +\! \dot{P}_{(p,v)}({\bf r})]\}.
\end{eqnarray}
Here    $$I^*_{m_i}\!({\bf r})\!\! =\!\![I_{H_{m_i}}\!({\bf r})\!+\!I_{P_{m_i}}\!({\bf r})V_{m_i}\!({\bf r})],$$
 $$I^*_p\!({\bf r})\!\!=\!\![I_{H_p}\!({\bf r})\!+\! V_p\!({\bf r})I_{P_p}\!({\bf r})]$$ and we have taken into account that  $\nabla\,( \beta_k({\bf r},t) V_k({\bf r},t)) \sim V_k (t)\nabla\beta_k({\bf r}, t)$.
%-----------------------------------------------------------------------------------------------------------------------------------

 Before going over to the macroscopic momentum balance equations, we should find a relation between the chemical potential and effective temperature of the magnon subsystem. From the quasi-equilibrium distribution $\rho_q(t)$  it follows that
 \begin{eqnarray}\label {d20}
\delta\lb N_{m_1}\!({\bf r})\rb\!\! =\!\!-\!\!\int\!\! d{\bf r}'\! \{\delta \beta_{m_1}\!({\bf r}'\!,\!t)\!(N_{m_1}\!({\bf r}),\! H_{m_1}\!({\bf r}'\!)\!) -\qquad\nonumber\\
 \beta\mu_{m_1}\!({\bf r}'\!,\!t)\!(N_{m_1}\!({\bf r}),\! N_{m_1}\!({\bf r}'\!)\!)\!\! -\!\beta_{m_1}V_{m_1}\!({\bf r}'\!,\!t)\!(N_{m_1}\!({\bf r}),\! P_{m_1}\!({\bf r}'\!)\!) \},\nonumber\\
 \end{eqnarray}
where
$$\delta\lb A\rb = \lb A\rb -\lb A\rb_0,\quad (A,B) =\int\limits_0^1 d\lambda Sp \{A\rho_0^\lambda\,\Delta B\,\rho_0^{1-\lambda}\}.$$
If one admits that $N_{m_1}({\bf r}) = const$  in a non-equilibrium but steady-state case, (\ref{d20}) implies that
\begin{eqnarray}\label {d21}
\mu_{m_1} \simeq (\beta_{m_1}/\beta-1)\,R- (\beta_{m_1}/\beta)\,R_1\nonumber\\
R= \frac{(N_{m_1}, H_{m_1})}{(N_{m_1}, N_{m_1})}\quad R_1 =  V_{m_1}\frac{(N_{m_1}, P_{m_1})}{(N_{m_1}, N_{m_1})}.
\end{eqnarray}
Note that as $\beta_{m_1}\rightarrow \beta,\,\, V_{m_1} =0$, the chemical potential of magnons tends to zero: $\mu_m \rightarrow 0$.
%-------------------------------------------------------------------------------------------------------------------------------------------------------------------------------------
\section*{\bf Macroscopic equations}

Inserting the entropy production operator (\ref{d22}) into the expression for the NSO (\ref{d17}), we average the operator equations (\ref{d14}) for momenta of the subsystems under discussion. Then we have
\begin{eqnarray}\label{d23}
\lb \dot{P}_{m_i}({\bf r})\rb^t\!=\qquad\qquad\qquad\qquad\qquad\qquad\nonumber\\
=\!\!-\!\!\! \int\limits_{-\infty}^0\!\!dt'\,e^{\epsilon t'}\!\!\!\int\!\! d{\bf r}'\! \{\beta(\nabla I_{P_{m_i}}\!({\bf r}),\!  I_{N_{m_j}}\!({\bf r}'\!,\!t'))\nabla\mu_{m_j}({\bf r}'\!,\!\bar{t})\! +\qquad\qquad\nonumber\\
+(\nabla I_{P_{m_i}}({\bf r}) ,\! I^*_{m_j}({\bf r}'\!,t'))\nabla\beta_{m_j}({\bf r}',\bar{t}) +\qquad\qquad\qquad\nonumber\\
 +( \dot{P}_{(m_i,v)}({\bf r}), \!\dot{P}_{(m_j,v)}({\bf r}'\!,t'))\,\beta V_{m_j} ({\bf r}',\bar{t}) \},\qquad\qquad\qquad
\end{eqnarray}
here $\bar{t}\equiv t+t'$. The first summand in the right-hand side of (\ref{d23}) describes the diffusion and drift of magnons due to the  gradients of the chemical potential and the temperature, the last summand  the magnon-magnon scattering processes both inside each of the magnon subsystems and the "coherent"\, magnon scattering by the "thermal"\, magnons.

Analogously, we characterize the relaxation processes in the phonon subsystem:
\begin{eqnarray}\label{d24}
\lb \dot{P}_p({\bf r})\rb^t =\qquad\qquad\qquad\qquad\qquad\nonumber\\
=\!\!\int\limits_{-\infty}^0\!\!dt'e^{\epsilon t'}\!\!\int\!\! d{\bf r}'\!  \{(\nabla I_{P_p}({\bf r}), \!I^*_p({\bf r}'\!,\!t'))\,\nabla \beta_p\!({\bf r}',\bar{t}) + \qquad\qquad\nonumber\\
- ( \dot{P}_{(p,v)}({\bf r}), \dot{P}_{(m_i,v)}({\bf r}',t')) \beta V_i({\bf r}',\bar{t}) -\qquad\qquad \nonumber\\
 ( \dot{P}_{(p,v)}({\bf r}), \dot{P}_{(p,v)}({\bf r}',t'))\, \beta V_p({\bf r}',\bar{t}) \}.\qquad\qquad \nonumber\\
\end{eqnarray}
Given that the chemical potential and the effective temperature are related as in (\ref{d21}),  we introduce the general diffusion coefficient $D_{m_im_j}({\bf r},{\bf r}',t')$ :
\begin{eqnarray}\label{d25}
\beta(\nabla I_{P_{m_i}}({\bf r}),\! I_{N_{m_j}}\!({\bf r}',t'))\nabla\mu_{m_j}\!({\bf r}'\!,\bar{t})\!+\nonumber\\
+\!(\nabla I_{P_{m_i}}\!({\bf r}) , I^*_{m_j}\!({\bf r}'\!,\!t'))\nabla\beta_{m_j}\!({\bf r}'\!,\!\bar{t})=\nonumber\\
=D_{m_im_j}\!({\bf r}\!,\!{\bf r}'\!,\!t') \nabla\mu_{m_j}\!({\bf r}'\!,\!\bar{t})
\end{eqnarray}
where
\begin{eqnarray}\label{d26}
 \beta D_{m_im_j}\!({\bf r}\!,{\bf r}'\!,\!t')\!=\!(\nabla I_{P_{m_i}({\bf r})},\! I_{N_{m_j}}\!({\bf r}',t'))\ +\nonumber\\
 +(\nabla I_{P_{m_i}}\!({\bf r})\! ,\! I^*_{m_j}\!({\bf r}'\!,\!t')\!)/(R-R_1).
\end{eqnarray}
The above equations represent the temperature gradient as a driving force. Therefore, the entropy operator involves the additional summands such as $\beta_i({\bf r},t)\,V_i$ instead of $\beta\,V_i({\bf r},t)$.
%----------------------------------------------------------------------------------------------------------------

Now, revealing explicitly the correlation functions describing the relaxation processes and appearing in the momentum balance equations (\ref{d23}),  (\ref{d24}), we have
\begin{eqnarray}\label{d27}
( \dot{P}_{(m_i,v)}({\bf r}),\! \dot{P}_{(m_j,v)}({\bf r}',t'))\! =\qquad\qquad\nonumber\\
=\! ( \dot{P}_{(m_i,mp)}({\bf r}),\! \dot{P}_{(m_i,mp)}({\bf r}',t')) \!+\qquad\qquad\nonumber\\
 +( \dot{P}_{(m_i,m_im_j)}({\bf r}), \!\dot{P}_{(m_i,m_im_j)}({\bf r}',t')) ,\qquad
 \end{eqnarray}
 The first summand in the right-hand side of (\ref{d27}) describes the magnon-phonon scattering, the second the magnon-magnon scattering processes both inside each of the magnon subsystems and the \,injected"\, magnon scattering by the \,"thermal"\, magnons.
 \begin{eqnarray}\label{d28}
( \dot{P}_{(p,v)}\!({\bf r}),\! \dot{P}_{(p,v)}({\bf r}'\!,t'))\, =\qquad\qquad\nonumber\\
=( \dot{P}_{(p,pp)}\!({\bf r}), \!\dot{P}_{(p,pp)}\!({\bf r}'\,,\,t')) \!+\qquad\qquad\qquad\nonumber\\
+\! ( \dot{P}_{(p,pm)}\,({\bf r}),\,\dot{P}_{(p,pm)}\!({\bf r}'\,,\!t')).\qquad
\end{eqnarray}
The first term describes the processes of non-magnon relaxation of phonons; the second one governs the magnon-phonon scattering.
%---------------------------------------------------
Introducing the notation
\begin{equation}\label{d28a}
L_{(k,v)}({\bf r},{\bf r}',t') =  ( \dot{P}_{(k,v)}({\bf r}), \dot{P}_{(k,v)}({\bf r}',t')),
 \end{equation}
we re-write down the momentum balance equations in a convenient form for further analysis
\begin{eqnarray}\label{d29}
\lb \dot{P}_{m_1}({\bf r})\rb^t=\qquad\qquad\qquad\nonumber\\
\!\!-\!\! \int\limits_{-\infty}^0\!\!dt'\,e^{\epsilon t'}\!\!\int\!\! d{\bf r}' \beta\{D_{m_1m_1}\!({\bf r}\!,\!{\bf r}'\!,\!t')\,\nabla\mu_{m_1}\!({\bf r}'\!,\!\bar{t})\ +\qquad\qquad\nonumber\\
+ L_{(m_1, m_1p)}({\bf r},{\bf r}', t')\,\delta V_{m_1, p}\!({\bf r}'\!,\!\bar{t}) +\qquad\qquad\nonumber\\
+ L_{(m_1, m_1m_2)}\!({\bf r}\!,\! {\bf r}'\!,\!t')\,\delta V_{m_1,m_2}\! ({\bf r}'\!,\!\bar{t}) \},\qquad
\end{eqnarray}
%-----------------------------------------------------------------------------------------------

\begin{eqnarray}\label{d30}
\lb \dot{P}_{m_2}({\bf r})\rb^t=\qquad\qquad\qquad\nonumber\\
\!\!-\!\! \int\limits_{-\infty}^0\!dt'\,e^{\epsilon t'}\!\!\int\!\! d{\bf r}' \beta\{D_{m_2m_2}({\bf r}\!,{\bf r}'\!t')\,\nabla\mu_{m_2}\!({\bf r}'\!,\bar{t})\
+\qquad\qquad\nonumber\\
+ L_{(m_2, m_2p)}({\bf r}\!,{\bf r}',\!t')\,\delta V_{m_2, p}({\bf r}'\!,\bar{t})
+\qquad\qquad\nonumber\\
+L_{(m_2, m_1m_2)}({\bf r}\!,{\bf r}',\!t')\,\delta V_{m_2,m_1} ({\bf r}'\!,\bar{t}) \},\qquad
\end{eqnarray}

\begin{eqnarray}\label{d31}
\lb \dot{P}_p({\bf r})\rb^t =\qquad\qquad\qquad\nonumber\\
\!\! -\!\!\int\limits_{-\infty}^0dt'\,e^{\epsilon t'}\!\int\!\! d{\bf r}' \beta\{- D_{pp}({\bf r}\!,{\bf r}',\!t')\,\nabla  \beta_p({\bf r}'\!,\bar{t})
+\qquad\qquad\nonumber\\
 + L_{(p, m_1p)}({\bf r},{\bf r}',t')\, \delta V_{p,m_1}({\bf r}',\bar{t})  +\qquad\qquad\nonumber\\
 +  L_{(p, m_2p)}({\bf r}\!,{\bf r}'\!,t')\, \delta V_{p,m_2}({\bf r}',\bar{t})+\qquad\qquad\nonumber\\
 +  L_{(p, pp)}({\bf r},{\bf r}',t') V_p({\bf r}'\!,\bar{t})\}.\qquad\qquad
\end{eqnarray}
Here $ \delta V_{ik}\! = \! V_i - V_k,$\,\, $  D_{pp}({\bf r}\!,{\bf r}',\!t') = \beta(\nabla I_{P_p}({\bf r}) , I^*_p({\bf r}',t')).$

Equations  (\ref{d29}) - (\ref{d31}) allow conducting the analysis of how the interaction between the    subsystems at hand  affects the implementation of the drag effect. We introduce the average values of the forces induced by the chemical potential and temperature gradients:
$$F_{m_i}({\bf r})\! =\! \int\limits_{-\infty}^0\!dt'\,e^{\epsilon t'}\!\int\!  d{\bf r}'\,D_{m_im_j}({\bf r}\!,{\bf r}'\!,t')\,\nabla\mu_{m_j}({\bf r}',\bar{t})$$
$$F_p({\bf r})\! =\! \int\limits_{-\infty}^0\!dt'\,e^{\epsilon t'}\!\int\!  d{\bf r}'  D_{pp}({\bf r},{\bf r}',t')\,\nabla  \beta_p({\bf r}',\bar{t}). $$
 Besides,  introduce the inverse times of the magnon and phonon momentum relaxation caused by interaction with phonons  processes of  non-magnon relaxation of phonons. Let us designate them as $\omega_{(mp)}$,  and $\omega_{(pp)}$, respectively \cite{38, 39}
\begin{eqnarray}\label{d34}
\omega_{(\gamma,  v)}= ( P_{\gamma}, P_{\gamma})^{-1} \int\limits_{-\infty}^0\,dt'\,e^{\epsilon t'}  ( \dot{P}_{(\gamma,v)},
 \dot{P}_{(\gamma,v)}(t')),\nonumber\\
 \qquad\qquad \gamma = m_1, m_2, p\qquad
 \end{eqnarray}
%==========================================
We restrict ourselves to the discussion of a stationary case. For this purpose, we average the balance equations over time $t$.
To start the analysis, we consider the simplest case when the drift velocities of the magnon systems are  equal:  $V_{m_1} = V_{m_2}\equiv V_m$ and $\beta_{m_1}=\beta_{m_2}$. This actually means that we deal with one magnon and one phonon systems. In addition, we shall assume that the phonon momentum is maintained from the outside by an unchanged. Then, the momentum balance equations appear as

\begin{eqnarray}\label{d32}
 F_{m} =\mathscr{P}_m\, \omega_{(m, mp)}\,( V_m -V_p),
\end{eqnarray}
\begin{eqnarray}\label{d33}
0 = \mathscr{P}_p\, \omega_{ (p, mp)}\, (V_p -V_m)   + \mathscr{P}_p\, \omega_{(p, pp)} V_p.
\end{eqnarray}
The balance equation for the magnon momentum acquires the form
\begin{eqnarray}\label{d35}
F_m = V_m \frac{ \omega_{(p, pp)}\,\omega_{(m, mp)}}{\omega_{(m, mp)} + \omega_{(p, pp)}}\,\mathscr{P}_m.
\end{eqnarray}
where  $\mathscr{P}_m\equiv ( P_m, P_m) ,\quad \mathscr{P}_p \equiv ( P_p, P_p)$.
Finally, the dragging leads to the change in frequency of the magnon-phonon collisions, and the quantity
$$\Omega = \frac{\omega_{(p, pp)}\,\omega_{(m, mp)}}{\omega_{(m, mp)} + \omega_{(p, pp)}}$$
is the inverse relaxation time of the magnon momentum by non-equilibrium phonons.

From the expression (\ref{d35}) it follows that the drag effect has an influence on the magnon-phonon collision frequency. The phonon subsystem almost always remains in equilibrium, and the inverse relaxation time is defined by the frequency $\omega_{(m, mp)}$  provided that the inequality $\omega_{(p, pp)}>\omega_{(m, pm)}$  is fulfilled. The latter means that the phonon momentum gained quickly relaxes in the processes of non-magnon relaxation. If the opposite inequality $\omega_{(p, pp)}<\omega_{(m, pm)}$ holds, the leakage of the phonon momentum occurs slower than the gain momentum rate in the magnon-phonon collisions. In this case, the mechanism of the non-magnon phonon relaxation mainly contributes to the drag effect. In addition,
\begin{eqnarray}\label{d36}
\bar{\omega}_{(m, mp)}\!\simeq \!\omega_{(p, pp)}\!(\omega_{(m, mp)}/\omega_{(m, pm)})\! =\nonumber\\
=\!\omega_{(p, pp)}\!(\mathscr{P}_p /\mathscr{P}_m)\!  =\!\mathscr{P}_m\!\int\limits_{-\infty}^0\,dt\, e^{\epsilon t}\!(\dot{P}_{(p, pp)},\dot{P}_{(p, pp)}).
\end{eqnarray}
Thus, the criterion of realizing the drag effect consists in the requirement $\omega_{(m, pm)}\! >\!\omega_{(p, pp)}$  that coincides with the solution of the kinetic equation \cite{39}. It is worth emphasizing that to calculate the correlation function in the formula for $\omega_{(p,pp)}$, it is necessary to know particular mechanisms of the non-magnon phonon momentum relaxation. For considering the drag effects there are two mechanisms such as the Herring mechanism $\omega_{(p, pp)}\sim(k_BT)^3$  and the Simons mechanism $\omega_{(p, pp)}\sim(k_BT)^4$ leading to a rather strong temperature dependence of the relaxation frequencies.

Another limiting case corresponds to the situation when the drift velocities of thermal magnons and phonons are equal to $V_{m_2} = V_p $ и $(\beta_{m_2}=\beta_p)$. In this case, thermal magnons and phonons form one subsystem. From balance equations (\ref{d29}), (\ref{d30}) we obtain
\begin{eqnarray}\label{d37}
F_{m_1} = V_{m_1} \frac{ \omega_{(p, pp)}\,[\omega_{(p, mp)} + \omega_{(m, m_1m_2)}]}{\omega_{(p, mp)} + \omega_{(p,pp)}}.
\end{eqnarray}
From the expression (\ref{d37}) it follows that if $\omega_{(p,pp)}\,\gg\,\omega_{(p,mp)}$ then $F_1\!\sim\!\omega_{(p,mp)}$ and $F_1\!\sim\! \omega_{(m,m_1m_2)}$ if $\omega_{(p,mp)}\ll\omega_{(m,m_1m_2)}.$   If the opposite inequality, when $\omega_{(p,pp)}\!\ll\!\omega_{(p,mp)}$ then $F_1\!\sim\!\omega_{(p, pp)}\,[1 + \omega_{(m,m _1m_2)}/\omega_{(p,mp)}\,]$ and $F_1\!\sim\!\omega_{(p, pp)}$ if $\omega_{(m,m_1m_2)}\!\ll\!\omega_{(p,mp)}$ and $F_1\!\sim\!\omega_{(p,pp)}\,\omega_{(m, m_1m_2)}/\omega_{(p,mp)}$ when $\omega_{(m,m_1m_2)}\!\gg\!\omega_{(p, mp)}$.
%================================================================================

Now we look into the drag effect in the event of two magnon and one phonon systems. Then, the momentum balance equations can be written as follows. The set of the equations (\ref{d32}), (\ref{d33}) implies
\begin{eqnarray}\label{d37}
F_{m_1}\! =\! \{ \omega_{(m_1,mp)}\!+\!\omega_{(m,m_1m_2)}\!-\!\frac{ \omega_{(m_1,mp)}\omega_{(m_1,mp)}}{\Omega}\}V_{m_1}\!-
\nonumber\\
-\!\{\frac{\omega_{(m_1,mp)}\omega_{(m_2,mp)}}{\Omega} \!+\!\omega_{(m,m_1m_2)}\}\cdot
\nonumber\\\cdot
\{\frac{F_{m_2}\!+\!(\omega_{(m,m_1m_2)}\! +\! \omega_{(m_1,mp)}\omega_{(m_2,mp)}/\Omega)V_{m_1}}
{ \omega_{(m_2,mp)}\!+\!\omega_{(m,m_1m_2)}\!-\!\omega_{(m_2,mp)}\omega_{(m_2,mp)}/\Omega}\},\qquad\qquad
\end{eqnarray}
where $\Omega =  \omega_{(m_1,mp)}+\omega_{(m_2,mp)} + \omega_{(p,pp)}.$

Let the energy transfer channels from the magnon subsystems to the phonon subsystem be equal
$\omega_{(m_1,mp)} = \omega_{(m_2, mp)}=\omega_{(m,mp)},\,\,V_{m_1}=V_m  $. In this case we have
\begin{eqnarray}\label{d38}
F_{m_1}\! =\! \{ \omega_{(m,mp)}\!+\!\omega_{(m,m_1m_2)}\!-\! \omega_{(m,mp)}/\Omega\}V_{m}\!
-\nonumber\\
-\!\{\omega_{(m,mp)}/\Omega\!+\!\omega_{(m,m_1m_2)}\}\times\nonumber\\
\times \{\frac{F_{m_2}\!+\!(\omega_{(m,m_1m_2)}\! +\! \omega_{(m,mp)}/\Omega)V_{m}}
{ \omega_{(m,mp)}\!+\!\omega_{(m,m_1m_2)}\!-\!\omega_{(m,mp)}/\Omega}\}.\qquad\quad
\end{eqnarray}
Here $\Omega = 2+ \omega_{(p,pp)}/\omega_{(m,mp)}$.
If $ \omega_{(p,pp)}\!\gg\!\omega_{(m,mp)} $, then
\begin{eqnarray}\label{d39}
F_{m_1}\! =\! \{ \omega_{(m,mp)}\!+\!\omega_{(m,m_1m_2)}\}V_{m}\!
- \nonumber\\
-\!\omega_{(m,m_1m_2)}\cdot
\{\frac{F_{m_2}\!+\!\omega_{(m,m_1m_2)}}{\omega_{(m,mp)}\!+\!\omega_{(m,m_1m_2)}}\}.
\end{eqnarray}
The expression (\ref{d39}) claims that the spin-wave current $\sim F_1$   is determined by the relations between the correlation functions $\omega_{(m,m_1m_2)}$ and $\omega_{(m,mp)}$. As it follows from the expression (\ref{d39}) that if $\omega_{(m,m_1m_2)}\!\ll\!\omega_{(m,mp)}$ then $F_1\!\sim\!\omega_{(m,mp)}$. In this case magnon-phonon interaction is the dominant channel of a magnon relaxation.   If we have the opposite inequality  $\omega_{(m,m_1m_2)}\!\gg\!\omega_{(m,mp)}$ then $F_1\!\sim\!\omega_{(m,m_1m_2)}$.  In this case the  interaction between "injected"\, and "thermal"\, magnons is the dominant channel of a magnon relaxation.  Moreover, the inelastic scattering of the\, "injected"\, magnons by "thermal"\, ones can be regarded as scattering by impurity centers whose concentration is  temperature-varied.   This interaction will determine the temperature-field behaviour of the spin-wave current under the conditions of the Seebeck spin effect.

Because of the existence of two relaxation channels (magnon-phonon and magnon-magnon), the inelastic scattering of the "injected"\, magnons by "thermal"\, ones may give rise to the bottleneck effect and heating of the "thermal"\, magnons. Such a situation emerges if the "thermal"-magnon subsystem gains energy through the magnon-magnon channel faster than loses it along the magnon-phonon channel, i.e. $\omega_{(m,m_1m_2)}\!\gg\!\omega_{(m,mp)}.$

\section*{\bf Conclusion}

The formation of the two: "injected"\, (УcoherentФ) and УthermallyФ excited, different in energies magnon subsystems and the influence of its interaction with phonons and between on drag effect under spin Seebeck effect conditions in the magnetic insulator part of the metal/ferromagnetic insulator/metal structure is studied. The analysis of the macroscopic momentum balance equations of the systems of interest conducted for different ratios of the drift velocities of the magnon and phonon currents show that the УinjectedФ\, magnons relaxation on the УthermalФ\, ones is possible to be dominant over its relaxation on the phonons.  This interaction will be the defining in the forming of the temperature dependence of the spin-wave current under SSE conditions, and inelastic part of the magnon-magnon interaction is the dominant spin relaxation mechanism.
The existence of the two relaxation channels (magnon-phonon and magnon-magnon) in the case of inelastic scattering of the УinjectedФ\, magnons on the УthermalФ\, ones is shown to be leading to the УwarmingФ\, of the letter and to Narrow-neck effect. Such situation could be realized in the case of energy input rate threw the magnon-magnon channel to the "thermal"\, magnons domination over its leaking rate threw magnon-phonon mechanism

{\bf Acknowledgments }

The given work has been done as the part of the state task on the theme "Electron" 01201463330 (project 12-T-2-1011) with the support of the Ministry of Education of the Russian Federation (Grant 14.Z50.31.0025)

\end{document}